\documentclass[%
twocolumn,
superscriptaddress,
 amsmath,amssymb,
 aps,
 pra,
floatfix,
]{revtex4}
\usepackage[normalem]{ulem}
\usepackage{cancel}
\usepackage{physics}
\usepackage{braket}
\usepackage{graphicx}
\usepackage{dcolumn}
\usepackage{bm}
\usepackage[breaklinks=true,colorlinks,citecolor=blue,linkcolor=blue,urlcolor=blue]{hyperref}

\newcommand{\figpanel}[2]{Fig.~\hyperref[#1]{\ref*{#1}(#2)}}

\newcommand{\be}{\begin{equation}}
\newcommand{\ee}{\end{equation}}
\newcommand{\bea}{\begin{eqnarray}}
\newcommand{\eea}{\end{eqnarray}}

\begin{document}

\title{A non-perturbative no-go theorem for photon condensation in approximate models}

\author{G.M. Andolina   }

\affiliation{NEST, Scuola Normale Superiore, I-56126 Pisa,~Italy}
\affiliation{Istituto Italiano di Tecnologia, Graphene Labs, Via Morego 30, I-16163 Genova,~Italy}
\affiliation{ICFO - Institut de Ciencies Fotoniques,
Mediterranean Technology Park, 08860 Castelldefels (Barcelona),~Spain}
\author{F.M.D. Pellegrino}
\affiliation{Dipartimento di Fisica e Astronomia ``Ettore Majorana'', Universit\`a di Catania, Via S. Sofia 64, I-95123 Catania,~Italy}
\affiliation{INFN, Sez.~Catania, I-95123 Catania,~Italy}
\author{A. Mercurio}
\email[]{alberto.mercurio@unime.it}
\affiliation{Dipartimento di Scienze Matematiche e Informatiche, Scienze Fisiche e  Scienze della Terra,
	Universit\`{a} di Messina, I-98166 Messina, Italy}
\author{O. Di Stefano}
	\affiliation{Dipartimento di Scienze Matematiche e Informatiche, Scienze Fisiche e  Scienze della Terra, Universit\`{a} di Messina, I-98166 Messina, Italy}
\author{M. Polini}
\affiliation{Dipartimento di Fisica dell'Universit\`a di Pisa, Largo Bruno Pontecorvo 3, I-56127 Pisa, Italy}
\affiliation{School of Physics \& Astronomy, University of Manchester, Oxford Road, Manchester M13 9PL, United Kingdom}
\affiliation{Istituto Italiano di Tecnologia, Graphene Labs, Via Morego 30, I-16163 Genova,~Italy}
\author{S. Savasta}
\affiliation{Dipartimento di Scienze Matematiche e Informatiche, Scienze Fisiche e  Scienze della Terra,
	Universit\`{a} di Messina, I-98166 Messina, Italy}

\date{\today}

\begin{abstract}
Equilibrium phase transitions between a normal and a photon condensate state (also known as superradiant phase transitions) are a highly debated research topic, where proposals for their occurrence and no-go theorems have chased each other for the past four decades.
Recent no-go theorems have demonstrated that gauge invariance forbids second-order phase transitions to a photon condensate state when the cavity-photon mode is assumed to be {\it spatially uniform}. However, it has been theoretically predicted that a collection of three-level systems coupled to light can display a first-order phase transition to a photon condensate state.
Here, we demonstrate a general no-go theorem valid also for truncated, gauge-invariant models which forbids first-order as well as second-order superradiant phase transitions in the absence of a coupling with a magnetic field. In particular, we explicitly consider the cases of interacting electrons in a lattice and $M$-level systems.
\end{abstract}

\maketitle


\section{Introduction}
\label{sect:intro}
The Dicke model~\cite{dicke_pr_1954} is a paradigmatic model in the theory of light-matter interactions~\cite{gross_pr_1982,cong_josaB_2016,kockum_naturereviewsphysics_2019,kirton19}. It describes a collection of ${N}$ identical two-level systems coherently coupled to the same bosonic mode $\hat{a}$, arising from the quantization of the electromagnetic field inside a cavity of volume $V$.
As the name says, it was firstly introduced by Robert H. Dicke~\cite{dicke_pr_1954}, with the aim of describing the ``emission of coherent radiation'' obtained by considering a ``radiating gas as a single quantum-mechanical system". He dubbed such process ``super-radiant emission''.

In the thermodynamic limit (${ N}\to \infty$, $V\to \infty$, with ${ N}/V={\rm constant}$) and when the light-matter coupling strength exceeds a critical value,  the Dicke model undergoes an equilibrium second-order thermal phase transition~\cite{hepp_lieb,wang_pra_1973} between a normal and a ``super-radiant'' phase. In the zero-temperature limit, the phase transition persists and corresponds to a quantum phase transition~\cite{emary_brandes,buzek_prl_2005}. 
The super-radiant phase is characterized by a macroscopic number of photons, $\langle \hat{a} \rangle \sim \sqrt{{N}}$, and by a macroscopic number of excitations in the matter sector. To avoid confusion with the Dicke non-equilibrium super-radiant emission~\cite{dicke_pr_1954}, we here follow Refs.~\cite{andolina_prb_2019, Andolina20} and dub the equilibrium super-radiant phase transition as ``photon condensation''.

In the Coulomb gauge, a  careful derivation of the Dicke model starting from a microscopic condensed-matter model with electronic degrees of freedom leads to an additional diamagnetic term~\cite{rzazewski_prl_1975}, proportional to $(\hat{a}+\hat{a}^\dagger)^2$, which is usually neglected by utilizing a  (wrong) ``weak-coupling argument". It was soon understood~\cite{rzazewski_prl_1975,Knight,rzazewski_1979} that such additional term is crucial to preserve the  gauge invariance property of the model. Only when both terms generated by the minimal coupling substitution 
$\hat{\bm p}\to \hat{\bm p}+e {\bm A}/c$, 
(i.e.~the paramagnetic light-matter coupling and the diamagnetic term) are retained, does one have a gauge-invariant theory satisfying the Thomas-Reiche-Kuhn (TRK) sum rule~\cite{Sakurai,Tufarelli15, Savasta2021TRK}.
The occurrence of photon condensation in such a generalized Dicke model is forbidden~\cite{rzazewski_prl_1975,rzazewski_prl_2006,nataf_naturecommun_2010}.

Despite its importance, the Dicke model is not exhaustive at all. In recent years, researchers have transcended it by studying interactions between matter degrees of freedom and quantized electromagnetic fields in a variety of other models and physical systems.
 Photon condensation has been predicted in many of these ``beyond-Dicke'' systems, including three-level systems~\cite{Hayn11,Baksic13}, graphene~\cite{hagenmuller_prl_2012}, ferroelectric materials~\cite{keeling_jpcm_2007}, superconducting circuits~\cite{nataf_naturecommun_2010,ciuti_prl_2012,Jaako16,Bamba16}, and strongly correlated (a.k.a.~quantum) materials~\cite{mazza_prl_2019}. 

A number of no-go theorems for photon condensation in a single-mode spatially-uniform cavity field have appeared in the literature~\cite{Gawedzki_pra_1981,Bamba_pra_2014,chirolli_prl_2012,pellegrino_prb_2014,viehmann_prl_2011,andolina_prb_2019}, showing that gauge invariance forbids photon condensation even in such ``beyond-Dicke'' systems.  
Often, theorems have been opposed by ``go theorems"~\cite{Vukics12,Stokes20,nataf_naturecommun_2010}.
Photon condensation remains a rather controversial theoretical topic. 

At present, the most recent no-go theorem is reported in Ref.~\cite{andolina_prb_2019}, where the authors showed that photon condensation is forbidden by gauge invariance for generic non-relativistic interacting electron systems coupled to a spatially-uniform cavity mode. The proof is based on linear response theory~\cite{Pines_and_Nozieres,Giuliani_and_Vignale} and uses the smallness of the order parameter $\alpha=\langle \hat{a} \rangle$.  It is therefore valid only for second-order phase transitions, where $\alpha \ll 1$ at the phase transition, and changes continuously.
It is by now clear that a natural path to overcome such theorem is to consider spatially-varying cavity fields~\cite{guerci,Basko19,Andolina20}. In these recent works, photon condensation has been shown to occur and is essentially a magneto-static instability~\cite{Basko19,Andolina20,guerci,Zueco20}. 
Apparently, another possibility to bypass the hypothesis of such theorem could be to consider a first-order phase transition~\cite{Hayn11,Baksic13}, where the order parameter $\alpha$ abruptly changes from zero (in the normal phase) to a finite value (in the photon condensate phase).
As a matter of fact, that first-order phase transitions were a valuable possibility to overcome the no-go theorem was first discussed some time ago~\cite{ciuti_prl_2012,ReplyViehmann}. In these works, an ensemble of three level systems coupled to single uniform mode undergoes to first-order phase transition. Indeed, according to Refs.~\cite{ciuti_prl_2012,ReplyViehmann} systems displaying first-order phase-transitions were thought as valuable candidates to realize photon condensation.

These results are, however, in contrast with a rather general no-go theorem presented already in 1978 \cite{Knight}. In this work, an ensemble of electrons in the presence of single-particle potentials and interacting with a uniform electromagnetic mode is considered and it is shown that superradiant phase transitions (of any order) to a photon condensate are forbidden. In this proof, no truncation is taken and the full infinite-dimensional Hilbert space is retained.
However, it is often impractical to deal with an exponentially large Hilbert space. Hence, when performing explicit calculations in atomic systems, or more generally, in many-body systems, approximate (truncated) models are customarily employed. However, it has been shown that such approximations can spoil gauge invariance \cite{DiStefano19}. Since no-go theorems are closely related to gauge-invariance, it is natural to conclude that the super-radiant phase transition that can be found in these approximate models (e.g. in the three-level systems discussed in Refs.~\cite{Hayn11,Baksic13}) is a fictitious effect due to the Hilbert space truncation.
Since, in the ultra strong coupling, these models fails in describing the correct ground state, it is important to find a systematic procedure to construct truncated models fulfilling gauge-invariance and free of spurious phase transition. It has been shown that, even for these approximate models, it is possible to build light-matter interactions which consistently satisfy the gauge-invariance principle \cite{DiStefano19,Garziano20,Savasta21}. Here, we extend the no-go theorem for photon condensation of Ref.~\cite{Knight} for gauge-invariant truncated models of light-matter interacting systems.
In particular, in the first part of this Article, we consider an interacting system of electrons roaming in free space and on a lattice and we extend the no-go theorem of Ref.~\cite{Knight} to such a system. In the second part of this Article, we present a no-go theorem for a generic $M$-level matter system interacting with a uniform electromagnetic field. On the basis of our new no-go theorem, we conclude that the first-order phase transition phenomenology discussed in the pioneering works~\cite{Hayn11,Baksic13} on three-level systems coupled to a cavity mode is incorrect. The reason is that, in these models, the light-matter interaction was not derived from an underlying gauge-invariant model. Conversely, an {\it ad hoc} diamagnetic term was added. Such addition, which was made to enforce the TRK sum-rule~\cite{Sakurai,Tufarelli15, Savasta2021TRK}, is not always sufficient to prevent a breakdown of gauge invariance. While enforcing the TRK sum-rule alone was a reasonable approach at the time that Refs.~\cite{Hayn11,Baksic13} were published, nowadays more refined techniques to enforce gauge invariance in systems with an arbitrary but finite number of levels have been developed~\cite{DiStefano19,Garziano20,Savasta21,Roman22} and applied to a few solid-state systems~\cite{Schiro20,Li20b,Li22}. Such methods can be viewed as an application of lattice gauge theory~\cite{Wiese}.
Here, we also employ these new tools to derive a fully gauge-invariant model describing $M$-level systems  coupled to a cavity mode. In accordance with the general theorem, such model does not display photon condensation.  As an example, we analyze in detail an ensemble of three-level systems.

Our Article is organized as following. 
In Sect.~\ref{sect:general theorem} we present a non-perturbative  no-go theorem for photon condensation in the second-quantization framework. We consider both the continuum case and the case of Hilbert-space truncation on a lattice.
In Sect.~\ref{sect: M Levels} we present a non-perturbative  no-go theorem for photon condensation, valid for a generic $M$-level matter system interacting with the electromagnetic field via the electric dipole moments. Finally, in Sect.~\ref{sect:summary_and_conclusions} we draw our main conclusions.
\begin{figure}[t]
    \centering
    \includegraphics[width=\linewidth]{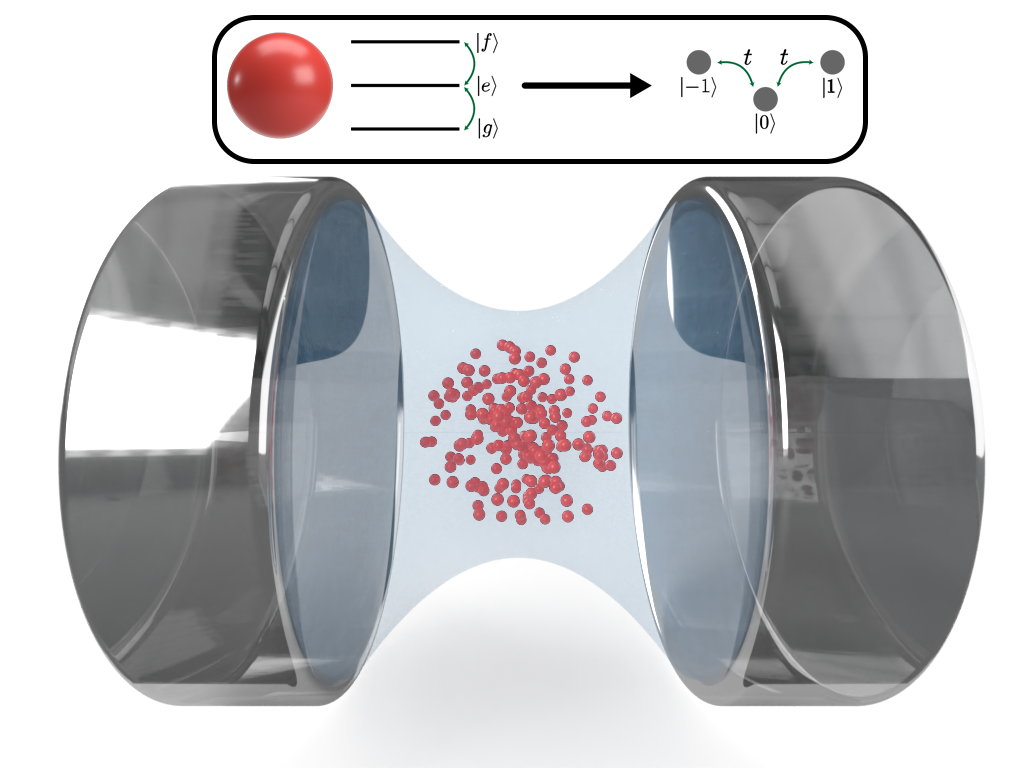}
    \caption{Pictorial representation of an ensemble of three-level atoms interacting with a cavity field. Each atom, which is considered in a ladder configuration, can be seen as a three-sites system with nearest neighbor hopping and parity symmetry. According to the lattice gauge theory, the interaction with the electromagnetic field can be obtained using the Wilson parallel transporter.}
    \label{fig:three level}
\end{figure}
\section{Gauge invariance, photon condensation, and no-go theorem in interacting electron systems}
\label{sect:general theorem}

\subsection{Interacting electron system in the continuum}
\label{sect:general theoremA}

We consider a quantum many-body system of interacting electrons, following the notation of Ref.~\cite{Schiro20}. In second quantization, the electronic Hamiltonian can be written as
\begin{equation}\label{eq:Hel}
\hat H_{\rm el} = \hat H_0 + \hat H_{\rm ee}~,    
\end{equation}
 where the one-body part, $\hat H_0$, reads as following 
\be
\label{eq: one-body electron Hamiltonian}
    \hat H_0 = \int d {\bm r} ~\hat \psi^\dag ({\bm r}) h_0({\bm r}) \hat \psi({\bm r})\, ,
\ee
with 
\be
h_0({\bm r}) = - \frac{{\hbar^2}\nabla^2}{2m} + V({\bm r})\, ,
\ee
while the electron-electron interaction contribution is given by
\be
\label{eq: ee-interact electron Hamiltonian}
\hat H_{\rm ee} = \int d{\bm r}\, d{\bm r}'\, 
\hat \psi^\dag ({\bm r})\,  \hat \psi^\dag ({\bm r}')\,
U(\abs{{\bm r} - {\bm r}'})\,
\hat \psi({\bm r}')\, \hat \psi({\bm r})\, .
\ee
Here, $V(\bm{r})$ and $U(\abs{{\bm r} - {\bm r}'})$ represent a generic one-body and two-body interaction potential, respectively.

The electron system is invariant under a global phase transformation $\hat \psi({\bm r}) \to e^{i \theta} \hat \psi({\bm r})$, and the associate Noether current reads
\be
 \hat {\bm J}({\bm r}) = \hat \psi^\dag ({\bm r})\, 
 (-i {\hbar}{\bm \nabla})\, \hat \psi ({\bm r}) + {\rm H. c.}\, .
\ee
However, the system is not invariant under a local phase transformation, $\hat \psi({\bm r}) \to e^{i \theta({\bm r})} \hat \psi({\bm r})$.
Such invariance can be restored by introducing the interaction with the electromagnetic field, by employing a minimal coupling scheme. Considering the Coulomb gauge---the effects of the scalar potential being already described by $V({\bm r})$ and $U(\abs{{\bm r} - {\bm r}'})$---the total light-matter Hamiltonian is given by:
\be\label{HC}
\hat H_{\rm C} = \int d{\bm r} 
\hat \psi^\dag ({\bm r})  h_{\rm C} ({\bm r})  \hat \psi({\bm r})
+ \hat H_{\rm ee} + \hat H_{\rm ph} \, ,
\ee
where
\be
 h_{\rm C}({\bm r}) = \hat T_{\bm A} + V({\bm r})\, ,
\ee
and 
\be
\hat T_{\bm A} = \frac{1}{2m}\left[-i {\hbar}{\bm \nabla} + \frac{e}{c} \hat {\bm A}({\bm r})\right]^2\, .
\ee
Here, $e>0$ is the elementary electron charge, $c$ is the speed of light, and $\hat {\bm A}({\bm r})$ is the space-dependent field operator vector describing the electromagnetic field in the Coulomb gauge. The Hamiltonian of the free field is given by
\be
 \hat H_{\rm ph} = { \frac{1}{8 \pi}} \int d{\bm r} \left\{ \hat {\bm \Pi}^2(\bm{r}) + [{\bm \nabla} \times \hat {\bm A}(\bm{r})]^2 \right\}\, ,
\ee
where $\hat {\bm \Pi}(\bm{r})$ is the conjugate momentum.

In this work, for simplicity, we will consider a single mode decomposition
of the fields \cite{Schiro20},
\bea
\hat {\bm A}({\bm r}) &=& {\bm A}_0({\bm r}) (\hat a + \hat a^\dag)\, ,  \\
\hat {\bm \Pi}({\bm r}) &=& i {\bm \Pi}_0({\bm r}) (\hat a - \hat a^\dag) \, ,
\eea
where $\int d \bm{r} ({\bm \nabla} \times{\bm A}_0(\bm{r}))^2/(2\pi)=\int d \bm{r} {\bm \Pi}^2_0(\bm{r})/(2\pi) =\hbar \omega_{\rm ph}$, where $\omega_{\rm ph}$ is the resonance frequency of the cavity mode, and ${\bm A}_0({\bm r}), {\bm \Pi}_0({\bm r})$ are the mode functions~\cite{Schiro20}.
 Notice that such single-mode approximation has been widely adopted in the Literature \cite{hepp_lieb,wang_pra_1973,emary_brandes,andolina_prb_2019,Hayn11,Baksic13,keeling_jpcm_2007,ciuti_prl_2012,viehmann_prl_2011,mazza_prl_2019,nataf_naturecommun_2010,guerci,Basko19,Zueco20,Roman22,Schiro20} in the context of photon condensation.

In terms of the single-mode photon creation ($\hat a^\dag$) and annihilation ($\hat a$) operators, the field Hamiltonian reduces to
\be
\label{eq: HPH}
\hat H_{\rm ph} ={\hbar} \omega_{\rm ph} \hat a^\dag \hat a\, .
\ee

A transformation of both the electronic and electromagnetic fields of the form
\bea
\hat \psi({\bm r}) \to e^{i \theta({\bm r})} \hat \psi({\bm r}) \\
\hat {\bm A} ({\bm r}) \to \hat {\bm A}({\bm r}) - \frac{\hbar c}{e} {\bm \nabla} \theta({\bm r})\, ,
\eea
leaves the Hamiltonian \eqref{HC} invariant, in agreement with the gauge principle.
We observe that Eq.~\eqref{HC} neglects the Zeeman coupling between the electron's spin and the magnetic component of the electromagnetic field. The absence of this term is justified either when the magnetic field is zero or when it can be neglected in the spatial region where the field interacts with the electron system, as, e.g., in the dipole approximation.

Whenever the interaction of the matter system with the magnetic field can be neglected, the vector potential entering the interaction terms can be locally expressed as the gradient of a scalar field
\be\label{Achi}
{\bm A}_0({\bm r})= {\bm \nabla} \chi({\bm r})\, .
\ee
In the dipole approximation, $\chi({\bm r})$  can be written as $\chi({\bm r}) = {\bf r} \cdot {\bf A}_0$, with ${\bf A}_0$ spatially uniform. 
Applying strictly the dipole approximation (uniform vector potential) to semiconductors implies a complete neglect of propagation effects inside the medium. In order to neglect the interaction of the electron system with the magnetic field, in
an extended system such as a semiconductor, one can
divide the whole medium into many cells of the same volume $V_{\rm cell}$ and apply the dipole approximation at each cell \cite{Cho1986, Savasta1996}.
The cell should be much smaller than the field wavelength (the usual choice is to take the unit cell of
the crystal as such a unit). Such partial relaxation of the dipole approximation (extended dipole approximation) can be realized using Eq.~\eqref{Achi}.
Notice that the dipole approximation has been customarily employed in the Literature \cite{hepp_lieb,wang_pra_1973,emary_brandes,andolina_prb_2019,keeling_jpcm_2007,ciuti_prl_2012,viehmann_prl_2011,mazza_prl_2019,nataf_naturecommun_2010,Schiro20}. We emphasize that Eq.~\eqref{Achi} implies the magnetic field $\bm B$  in the spatial region where the electronic field is non-negligible is zero, i.e., ~${\bm B}({\bm r})\equiv{\bm \nabla} \times {\bm A}_0({\bm r})={\bm \nabla}  \times {\bm \nabla} \chi({\bm r})=0$.

When the interaction of the matter system with the magnetic field can be neglected, the minimal coupling replacement can also be implemented by applying a unitary transformation to the bare electronic Hamiltonian. The unitary operator transforms the electronic  field operators as follows~\cite{Schiro20}:
\be
\label{eq: dressing electronic fields}
\hat {\cal U} \hat \psi ({\bm r})\hat {\cal U}^\dag = e^{i({ e/c\hbar}) \chi({\bm r})(\hat a + \hat a^\dag)} \hat \psi ({\bm r})\, ,
\ee
where 
\be 
\label{eq:transf}
\hat {\cal U} = \exp\left[
-i { \frac{e}{c\hbar}} (\hat a + \hat a^\dag)
\int d {\bm r}\, \hat \psi^\dag({\bm r})
\chi({\bm r}) \hat \psi({\bm r})
\right] \, .
\ee
Equation \eqref{eq: dressing electronic fields} is demonstrated in Appendix \ref{app: Derivation dressed electron fields}.

We stress that only the electronic Hamiltonian  has to be transformed applying the unitary operator in Eq.~\eqref{eq:transf}, while the photonic field $\hat a$ is unchanged.
The Hamiltonian in Eq.~\eqref{HC} can be rewritten as 
\bea\label{HCU}
 \hat H_{\rm C} &=& \hat H_{\rm ph} + \hat {\cal U} (\hat H_0 + \hat H_{\rm ee}) \hat {\cal U}^\dag ~.
\eea
In principle Eq.~(\ref{HCU}) could further simplified by noticing that the unitary transformation $\hat {\cal U}$ does not alter the electron-electron interaction contribution to the Hamiltonian \cite{Schiro20}, $\hat {\cal U} \hat H_{\rm ee} \hat {\cal U}^\dag =\hat H_{\rm ee} $. However, we do not need to employ this property for the sake of this proof. In Coulomb gauge, $\hat H_{\rm C}$ is the total Hamiltonian describing both light, matter and their interactions.

We now show a no-go theorem for photon condensation, by proving that the photonic operator cannot have a non-vanishing expectation value in the ground state, i.e. the super-radiant order parameter is zero $\braket{\hat{a}}$. In what follows we exhibit a proof by contradiction, showing that if 
 there exists a ground state $\ket{\psi_0}$ characterised by a non-zero super-radiant order parameter $\alpha\equiv\mel{\psi_0}{\hat{a}}{\psi_0}\neq 0$, then it is possible to find another state $\ket{\psi}$ with lower energy, in contrast to the hypothesis of $\ket{\psi_0}$ being the ground state. Specifically, we extend to the second quantization framework, a procedure that has been developed for first-quantization \cite{Knight}. Notice that several theoretical analyses of photon condensation \cite{nataf_naturecommun_2010,guerci,Zueco20,Andolina20}, including those that predicts its occurrence, neglect light-matter entanglement and assume that the system's ground state is factorized into matter and light wave-functions, i.e.~it uses a mean-field approximation for the light-matter interaction. Here we do not need to invoke this assumption.
Let us consider the following unitary operator,
\begin{equation}
    \label{eq: sec_quant no-go unitary op}
    \hat{\mathcal{T}} = \hat{\mathcal{D}}(\alpha) \exp \left[ -i \frac{2 e}{c \hbar} \Re (\alpha) \int d {\bm r} \hat{\psi}^\dag ({\bm r}) \chi({\bm r}) \hat{\psi} ({\bm r}) \right]   ~,
\end{equation}
where $\hat{\mathcal{D}}(\alpha)=\exp{- \alpha^* \hat{a} + \alpha \hat{a}^\dag} $ is the displacement operator characterized by a displacement $\alpha$. Photonic operators transform under the displacement as,
\begin{eqnarray}
\label{eq:displacement}
\hat{\mathcal{D}}(\alpha) \hat{a} \hat{\mathcal{D}}(\alpha)^\dag= \hat{a} - \alpha~,
\end{eqnarray}
The electronic and photonic fields transform under $\hat{\mathcal{T}}$ as,
\begin{eqnarray}
\label{eq: shift of nogo oper 1}
    \hat{\mathcal{T}} \hat{\psi} ({\bm r}) \hat{\mathcal{T}}^\dag &=& e^{i \frac{2 e}{c \hbar} \Re (\alpha) \chi({\bm r})} \hat{\psi}({\bm r})~ \\
\label{eq: shift of nogo oper 2}
\hat{\mathcal{T}} \hat{a} \hat{\mathcal{T}}^\dag &=& \hat{a} - \alpha~,
\end{eqnarray}
where in the second line we used that $\exp \left[ -i \frac{2 e}{c \hbar} \Re (\alpha) \int d {\bm r} \hat{\psi}^\dag ({\bm r}) \chi({\bm r}) \hat{\psi} ({\bm r}) \right] $  does not act on the photonic sector and then Eq.~\eqref{eq:displacement} to transform the photon operator.

We remind that we assumed as an hypothesis that the Hamiltonian of Eq.~\eqref{HCU} has a ground state $\ket{\psi_0}$ with a non-vanishing expectation value of the photonic annihilation operator ($\alpha \equiv \mel{\psi_0}{\hat{a}}{\psi_0} \neq 0$). We now consider the state $\ket{\psi} = \hat{\mathcal{T}}^\dag \ket{\psi_0}$. By means of Eq.~\eqref{eq: shift of nogo oper 2} we can prove that such state has zero order parameter,
\begin{eqnarray}
\mel{\psi}{\hat{a}}{\psi}=\braket{\psi_0|\hat{\mathcal{T}}\hat{a}\hat{\mathcal{T}}^\dagger|\psi_0}=0,
\end{eqnarray}
where we employed the assumption $\braket{\psi_0|\hat{a}|\psi_0}=\alpha$.
In the following we show that the trial state  $\ket{\psi} = \hat{\mathcal{T}}^\dag \ket{\psi_0}$ has lower energy than $\ket{\psi_0}$, contradicting the initial assumption that $\ket{\psi_0}$ is the ground state.

First, we can prove that,
\begin{eqnarray}
    \label{eq: DUD}
{\hat{\mathcal{D}}}(\alpha) \hat{\mathcal{U}}^\dagger \hat{\mathcal{D}}(\alpha)^\dag=\exp \left[ -i \frac{2 e}{c \hbar} \Re (\alpha) \int d {\bm r} \hat{\psi}^\dag ({\bm r}) \chi({\bm r}) \hat{\psi} ({\bm r}) \right]  \hat{\mathcal{U}}^\dagger~,\nonumber\\
\end{eqnarray}
where we used that, by means of Eq.~\eqref{eq: shift of nogo oper 2}, $ \hat{\mathcal{D}}(\alpha)(\hat{a}+\hat{a}^\dagger) \hat{\mathcal{D}}(\alpha)^\dag=(\hat{a}+\hat{a}^\dagger)-2\Re (\alpha)$.
Before proceeding, it is useful to consider the operators product $\hat{\mathcal{U}}^\dagger\hat{\mathcal{T}}^\dagger$,
\begin{eqnarray}
    \label{eq: TU0}
\hat{\mathcal{U}}^\dagger\hat{\mathcal{T}}^\dagger&=&\hat{\mathcal{D}}^\dagger(\alpha)\hat{\mathcal{D}}(\alpha)\hat{\mathcal{U}}^\dagger\hat{\mathcal{D}}^\dagger(\alpha) \nonumber \times \\&\times& \exp \left[ i \frac{2 e}{c \hbar} \Re (\alpha) \int d {\bm r} \hat{\psi}^\dag ({\bm r}) \chi({\bm r}) \hat{\psi} ({\bm r}) \right] ~,\nonumber\\
\end{eqnarray}
where we expressed $\hat{\mathcal{T}}$ by using the definition in Eq.~\eqref{eq: sec_quant no-go unitary op} and we inserted a product of displacement operators by means of the identity $\hat{\mathcal{D}}^\dagger(\alpha)\hat{\mathcal{D}}(\alpha)=1$. 
By using Eq.~\eqref{eq: DUD} the previous expression can be simplified as,
\begin{eqnarray}
    \label{eq: TU}
\hat{\mathcal{U}}^\dagger\hat{\mathcal{T}}^\dagger=\hat{\mathcal{D}}^\dagger(\alpha)\hat{\mathcal{U}}^\dagger~.
\end{eqnarray}
Now we evaluate the total Coulomb Hamiltonian on the trial state $\ket{\psi}=\hat{\mathcal{T}}^\dagger\ket{\psi_0}$,
\begin{eqnarray}
    \label{eq: HcOverPsi}
 \braket{\psi|\hat{H}_C|\psi}=  \braket{\psi_0|\Big[ \hat{\mathcal{T}}\hat{\mathcal{U}}  (\hat H_0 + \hat H_{\rm ee}) \hat{\mathcal{U}}^\dagger \hat{\mathcal{T}}^\dagger+\hat{\mathcal{T}}\hat H_{\rm ph} \hat{\mathcal{T}}^\dagger \Big]|\psi_0}~.\nonumber\\
\end{eqnarray}
On one hand, the matter Hamiltonian can be simplified as, 
\begin{eqnarray}
 \label{eq: HmOverPsi}
 \braket{\psi_0| \hat{\mathcal{T}}\hat{\mathcal{U}} (\hat H_0 + \hat H_{\rm ee}) \hat{\mathcal{U}}^\dagger \hat{\mathcal{T}}^\dagger|\psi_0}=  \braket{\psi_0| \hat{\mathcal{U}}^\dagger  (\hat H_0 + \hat H_{\rm ee}) \hat{\mathcal{U}} |\psi_0}~,\nonumber\\
\end{eqnarray}
where we used the property $\hat{\mathcal{U}}^\dagger\hat{\mathcal{T}}^\dagger= \hat{\mathcal{D}}^\dagger(\alpha) \hat{\mathcal{U}}^\dagger$ given in Eq.~\eqref{eq: TU} and the fact that the displacement operator leaves invariant the matter Hamiltonian $\hat H_0 + \hat H_{\rm ee}$, $\hat{\mathcal{D}}(\alpha) (\hat H_0 + \hat H_{\rm ee})\hat{\mathcal{D}}^\dagger(\alpha) =\hat H_0 + \hat H_{\rm ee}$.
On the other hand, by means of Eq.~\eqref{eq: shift of nogo oper 2}, we can calculate the average value of the photonic Hamiltonian $\hat H_{\rm ph}$,
\begin{eqnarray}
    \label{eq: HphOverPsi}
 \braket{\psi_0|\hat{\mathcal{T}}\hat{H}_{\rm ph}\hat{\mathcal{T}}^\dagger|\psi_0}&=& \langle\psi_0|\Big[\hbar \omega_{\rm ph} ( \hat{a}^\dagger\hat{a}+\abs{\alpha}^2  )+\nonumber\\
 &-&  \hbar \omega_{\rm ph} (\alpha \hat{a}^\dag +\alpha^* \hat{a} )\Big] |\psi_0\rangle ~.
\end{eqnarray}
By using that, by construction, we have $\braket{\psi_0|\hat{a}    |\psi_0}=\alpha$, Eq.~\eqref{eq: HphOverPsi} simplifies to,
\begin{eqnarray}
    \label{eq: HphOverPsi1}
 \braket{\psi_0|\hat{\mathcal{T}}\hat{H}_{\rm ph}\hat{\mathcal{T}}^\dagger|\psi_0}=  \hbar \omega_{\rm ph}\big(\hat{a}^\dagger\hat{a}-|\alpha|^2\big) ~.\nonumber\\
\end{eqnarray}
By combining Eq.~\eqref{eq: HphOverPsi1} and Eq.~\eqref{eq: HmOverPsi} and the definition of the total Coulomb Hamiltonian Eq.~\eqref{HCU} we have, 
\begin{eqnarray}
    \label{eq: HcOverPsi1}
 \braket{\psi|\hat{H}_C|\psi}=\braket{\psi_0|\hat{H}_C|\psi_0}  -\hbar \omega_{\rm ph}|\alpha|^2 ~.
\end{eqnarray}
Noticing that $\hbar \omega_{\rm ph}|\alpha|^2$ is by hypothesis a positive and strictly non-zero quantity we have,
\begin{eqnarray}
 \braket{\psi|\hat{H}_C|\psi}<\braket{\psi_0|\hat{H}_C|\psi_0}~.
 \end{eqnarray}
This equation implies that the state $\ket{\psi_0}$, which has a non-vanishing expectation value of the photon annihilation operator $\hat{a}$, is not the real ground state of the system, since the state $\ket{\psi}$, which was built specifically to have a vanishing expectation value, has a lower energy. This concludes the proof by contradiction that super-radiant phase transitions to a photon condensate is forbidden for any interacting light-matter system which can be described by an effective Hamiltonian as Eq.~\eqref{HCU}.

We close by noticing that this result applies also to the case of a multi-mode cavity field, provided that it still corresponds to the physical situation of ${\bm B}={\bm 0}$.  In the absence of a magnetic field, the most general coupling to a transverse electric field is given by the following unitary transformation, 
\begin{eqnarray}
\hat {\cal U} = \exp  \left[
-i \frac{e}{c\hbar}\sum_i (\hat a_i + \hat a_i^\dag)
\int d {\bm r}\, \hat \psi^\dag({\bm r})
\chi_i({\bm r}) \hat \psi({\bm r})
 \right]~,
 \end{eqnarray}
where the index $i$ labels the different modes. The previous proof is generalized to the multi-mode case in Appendix \ref{app:multi-mode}.

While Eq.~\eqref{HC} neglects the Zeeman coupling, our main conclusion can be easily generalized also to the case in which such coupling is present. The Zeeman coupling is proportional to the scalar product of the electron spin operator and the magnetic field, i.e.~$ \hat{\bm{\sigma}}\cdot \hat{\bm{B}} (\bm{r})$.  Since in this work $\bm{B} (\bm{r})={\bm \nabla} \times {\bm A}_0({\bm r}) = {\bm 0}$, the Zeeman coupling does not alter the above analysis. 

Finally, we stress that the photon condensate order parameter has been defined as $\braket{\hat{a}}$ in the Coulomb gauge. The quantity $\braket{\hat{a}}$ is not a physical, gauge-invariant quantity~\cite{Vukics12,Stokes20,Settineri}. For example, in the dipolar gauge, $\braket{\hat{a}}$ measures spontaneous polarization of matter, which is a signature of ferroelectricity~\cite{keeling_jpcm_2007}. 
In contrast, we here choose as order parameter the displacement field due to transverse photons, which coincides with $\braket{\hat{a}}$ only in the Coulomb gauge. This is a well defined gauge-invariant quantity and our no-go theorem manifests in other gauges as the absence of a transverse electromagnetic field. Of course, when applying a gauge (unitary) transformation, invariant expectation values are obtined only transforming accordingly both the quantum states and the operators, see, e.g., Ref. \onlinecite{Settineri}.
{ In an arbitrary gauge, our main result should be read as follows, at equilibrium a transverse field cannot emerge spontaneously in a region where the magnetic field is absent.}

\subsection{Interacting electron system on a lattice}
The procedure discussed in the previous Subsection can be applied also to the case of interacting electron systems on a lattice, where the space domain is discretized. The main differences from the continuous case are: (i) the replacement of the integral with a discrete summation ($\int d{\bm r} \to \sum_{\bm r_i}$), (ii) the replacement of the electronic field with a fermionic annihilation operator ($\hat{\psi}({\bm r}) \to \hat{c}_{\bm r_i}$, with the anti-commutation property $\{\hat{c}_{\bm r_i}, \hat{c}_{{\bm r_j}}^\dag\} = \delta_{{\bm r_i}, {\bm r_j}}$). Hence, the one-body electron and the electron-electron interaction Hamiltonians, expressed in Eq.~\eqref{eq: one-body electron Hamiltonian} and \eqref{eq: ee-interact electron Hamiltonian} respectively, become
\bea
\label{eq: lattice one-body electron hamiltonian}
\hat{H}_0 &=& \sum_{\bm{r}_i, \bm{r}_j} \tau_{\bm{r}_i, \bm{r}_j} \hat{c}_{\bm r_i}^\dag \hat{c}_{\bm r_j}\\
\label{eq: lattice ee-interact hamiltonian}
\hat{H}_{\rm ee} &=& \sum_{\bm{r}_i, \bm{r}_j} U_{\bm{r}_i, \bm{r}_j} \hat{c}_{\bm r_i}^\dag \hat{c}_{\bm r_j}^\dag \hat{c}_{\bm r_j} \hat{c}_{\bm r_i}\, ,
\eea
where $\tau_{\bm{r}_i, \bm{r}_j} = \Omega_{\bm r_i} \delta_{\bm{r}_i, \bm{r}_j} + t_{\bm r_i} \delta_{\langle \bm{r}_i, \bm{r}_{j} \rangle}$ describes the on-site energies and the near-neighbor hopping factors (where $\langle \bm{r}_i, \bm{r}_{j} \rangle $ denotes near-neighbor sites), while $U_{\bm{r}_i, \bm{r}_j}$ is a symmetric operator, since the electron-electron interaction potential $U(\abs{\bm{r} - \bm{r}'})$ expressed in Eq.~\eqref{eq: ee-interact electron Hamiltonian} depends only on the distance between the two points $\bm r$ and $\bm r'$.

However, in general, the truncation of the Hilbert space, could introduce some kind of spatial non-locality in the electron-electron interaction \cite{Ismail2001}. Hence, it can be useful to also consider the generalized version of $\hat{H}_{\rm ee}$ which includes also non-local effects
\be
\label{eq: lattice non-local ee-interact hamiltonian}
\hat{H}_{\rm ee}^{\rm nl} = \sum_{\substack{ \bm{r}_i, \bm{r}_j \\ \bm{r}_{l}, \bm{r}_{m} }} U_{\bm{r}_i, \bm{r}_j}^{\bm{r}_l, \bm{r}_m} \hat{c}_{\bm r_i}^\dag \hat{c}_{\bm r_j}^\dag \hat{c}_{\bm r_m} \hat{c}_{\bm r_l}\, .
\ee
The previous Hamiltonian appears for example in the context of non-Fermi liquid states of matter. With a suitable choice of the parameters $U_{\bm{r}_i, \bm{r}_j}^{\bm{r}_l, \bm{r}_m}$ it indeed coincides with the so-called SYK model \cite{Sachdev93,Kitaev15,gu_2019}.
The generalized electron-electron interaction term $\hat{H}_{\rm ee}^{\rm nl}$ reduces to the usual interaction Hamiltonian $\hat{H}_{\rm ee}$ for
\[
 U_{\bm{r}_i, \bm{r}_j}^{\bm{r}_l, \bm{r}_m} =
 U_{\bm{r}_i, \bm{r}_j} \delta_{\bm{r}_i, \bm{r}_l}
 \delta_{\bm{r}_j, \bm{r}_m}\, .
\]

The interaction with a single-mode cavity field is again introduced by applying a unitary transformation to the electronic fields (which now become the fermionic operators $\hat{c}_{\bm r_i}$) in a manner similar to Eq.~\eqref{eq: dressing electronic fields}, that is
\be
\label{eq: dressed lattice field}
\mathcal{\hat{U}} \hat{c}_{\bm r_i} \hat{\mathcal{U}}^\dag = e^{i(e / c \hbar) \chi_{\bm r_i} (\hat{a} + \hat{a}^\dag)} \hat{c}_{\bm r_i}\, , 
\ee
with
\be
\hat{\mathcal{U}} = \exp \left[ -i \frac{e}{c \hbar} (\hat{a} + \hat{a}^\dag) \sum_{\bm r_i} \chi_{\bm r_i} \hat{c}_{\bm r_i}^\dag \hat{c}_{\bm r_i} \right]\, .
\ee
Eq.~\eqref{eq: dressed lattice field}, which is demonstrated in Appendix \ref{app: Derivation dressed electron fields}, can be seen as the equivalent of the Peierls substitution \cite{Peierls1933}. Such procedure can be regarded as a particular instance of lattice gauge theory,  the general method developed by Wilson for studying non-perturbative relativistic gauge theories on a lattice \cite{Wilson}.
The obtained coupled light-matter Hamiltonian is similar to the continuum case expressed in Eq.~\eqref{HCU},
\be
\label{eq: lattice HCU}
\hat{H}_{\rm C} = \hat{H}_{\rm ph} + \hat{\mathcal{U}} \left( \hat{H}_0 + \hat{H}_{ee}^{\rm nl} \right) \hat{\mathcal{U}}^\dag\, ,
\ee
where $\hat{H}_{\rm ph}$ is given in Eq.~\eqref{eq: HPH} and represents the bare photonic Hamiltonian. Notice that, since we considered the generalized version of the electron-electron interaction term including non-locality, $\hat{H}_{\rm ee}^{\rm nl}$ may not commute with $\hat{\mathcal{U}}$ anymore. Nevertheless, this property is not needed for the sake of the proof, which holds also in the present case. However, we observe that the presence of such a nonlocal potential implies that the resulting total light-matter Hamiltonian will include additional terms arising from $\hat{\mathcal{U}} \hat{H}_{ee}^{\rm nl}  \hat{\mathcal{U}}^\dag$. These terms are crucial to ensure gauge invariance even in the presence of an effective non-local potential \cite{DiStefano19}.

The proof of the no-go theorem for interacting electrons systems on a lattice is now straightforward, and it follows the same steps applied to the continuum case in the previous Subsection. We start by introducing the lattice version of the unitary operator expressed in Eq.~\eqref{eq: sec_quant no-go unitary op}
\be
\hat{\mathcal{T}} = \exp \left[ -i  \frac{e}{c \hbar} 2 \Re (\alpha)   \sum_{\bm r_i} \chi_{\bm r_i} \hat{c}_{\bm r_i}^\dag \hat{c}_{\bm r_i} \right] \hat{\mathcal{D}}(\alpha)~,
\ee

which transforms the electronic and photonic operators as
\bea
\hat{\mathcal{T}} \hat{c}_{\bm r_i} \hat{\mathcal{T}}^\dag &=& e^{i {e}/({c \hbar})2 \Re (\alpha) \chi_{\bm{r}_i} } \hat{c}_{\bm r_i}~, \\
\hat{\mathcal{T}} \hat{a} \hat{\mathcal{T}}^\dag &=& \hat{a} - \alpha\, \label{eq:a_shift} .
\eea
Once again, we now suppose that the system described by the Hamiltonian \eqref{eq: lattice HCU} has a ground state $\ket{\psi_0}$ with a non-vanishing expectation value of the photonic annihilation operator. We now construct a trial state $\ket{\psi} = \hat{\mathcal{T}}^\dag \ket{\psi_0}$ with the property $\mel{\psi}{\hat{a}}{\psi} = 0$.
Following similar steps of the previous Subsection we can prove the property, $\hat{\mathcal{U}}^\dagger\hat{\mathcal{T}}^\dagger=\hat{\mathcal{D}}^\dagger(\alpha)\hat{\mathcal{U}}^\dagger$, corresponding to Eq.~\eqref{eq: TU}. It is useful to note that,
\begin{eqnarray}
    \label{eq: HcOverPsiII intro}
\hat{\mathcal{T}}\hat{\mathcal{U}}  (\hat H_0 + \hat H_{\rm ee}) \hat{\mathcal{U}}^\dagger \hat{\mathcal{T}}^\dagger&=&
\hat{\mathcal{U}} \hat{\mathcal{D}}(\alpha) (\hat H_0 + \hat H_{\rm ee})\hat{\mathcal{D}}^\dagger(\alpha) \hat{\mathcal{U}}^\dagger ~,\nonumber\\&=&
\hat{\mathcal{U}}  (\hat H_0 + \hat H_{\rm ee}) \hat{\mathcal{U}}^\dagger \,,
\end{eqnarray}
where  we used Eq.~\eqref{eq: TU} and the fact that $\hat{\mathcal{D}}(\alpha)$ commutes with $ \hat H_0 + \hat H_{\rm ee}$.
Hence, the total energy of the the trial state $\ket{\psi}$ reads:
\begin{eqnarray}
    \label{eq: HcOverPsiII}
 \braket{\psi|\hat{H}_C|\psi}= \braket{\psi_0|\Big[ \hat{\mathcal{U}}  (\hat H_0 + \hat H_{\rm ee}) \hat{\mathcal{U}}^\dagger +\hat{\mathcal{T}}\hat H_{\rm ph} \hat{\mathcal{T}}^\dagger \Big]|\psi_0}~.
\end{eqnarray}
%

From Eq.~\eqref{eq: HcOverPsiII} and employing Eq.~\eqref{eq:a_shift}, the energy finally reads,
\begin{eqnarray}
    \label{eq: HcOverPsi1II}
 \braket{\psi|\hat{H}_C|\psi}=\braket{\psi_0|\hat{H}_C|\psi_0}  -\hbar \omega_{\rm ph}|\alpha|^2 ~.
\end{eqnarray}

Again, we find that $\ket{\psi_0}$ cannot be the ground state of the system, since there is a lower energy state $\ket{\psi}$ with the property that $\mel{\psi}{\hat{a}}{\psi} = 0$, forbidding the superradiant phase transition for such system.

As we have seen, the presence of approximations, such as the discretization of the continuous space into a lattice, could introduce some kind of spatial non-locality. In addition, in solid-state physics, the transition from the continuum to the lattice is usually carried out in a slightly different way.  For example, according to the tight-binding approach, it is possible to have a number of orbitals on each lattice site. Following Ref.~\cite{Schiro20}, we can introduce the orbital index $\mu$ to each tight-binding site. In this case, then the one-body electron and the electron-electron non-local Hamiltonians become respectively
\bea
\hat{H}_0 &=& \sum_{\bm{r}_i, \bm{r}_j} \sum_{\mu_1, \mu_2} \tau_{\bm{r}_i, \bm{r}_j, \mu_1, \mu_2} \hat{c}_{\bm{r}_i, \mu_1}^\dag \hat{c}_{\bm{r}_j, \mu_2} \\
\hat{H}_{\rm ee}^{\rm nl} &=& \sum_{\substack{ \bm{r}_i, \bm{r}_j \\ \bm{r}_{l}, \bm{r}_{m} }} \sum_{\substack{ \mu_1, \mu_2 \\ \mu_3, \mu_3 }} U_{\bm{r}_i, \bm{r}_j, \mu_1, \mu_2}^{\bm{r}_l, \bm{r}_m, \mu_3, \mu_4} \hat{c}_{\bm r_i, \mu_1}^\dag \hat{c}_{\bm r_j, \mu_2}^\dag \hat{c}_{\bm r_m, \mu_3} \hat{c}_{\bm r_l, \mu_4}~, \nonumber
\eea
and the unitary operator $\hat{\mathcal{U}}$ becomes
\be
\hat{\mathcal{U}} = \exp \left[ -i \frac{e}{c \hbar} (\hat{a} + \hat{a}^\dag) \sum_{\bm{r}_i} \sum_{\mu} \chi_{\bm{r}_i, \mu} \ \hat{c}_{\bm{r}_i, \mu}^\dag \hat{c}_{\bm{r}_i, \mu} \right]~.
\ee
The proof of the no-go theorem follows the same procedure applied to the previous two cases.

\section{Gauge invariance, photon condensation, and no-go theorem in $M$-level systems}
\label{sect: M Levels}

In this Section we firstly generalize (Sect.~\ref{sec: No Go for m-level}) the no-go theorem to a truncated model composed by $M$-levels atoms showing that it does not display a transition to a photon condensate state (when placed in a spatially-uniform cavity field ${\bm A}$).

In order to derive a fully gauge invariant model for a system of three-level atoms interacting with a spatially-uniform cavity field, we show (Sect.~\ref{app:LatticeBasis}) that a generic $M$-level system can be mapped into a tight-binding model on a lattice with $M$ sites. In the third part of this Section (Sect.~\ref{subsect:TLA}), we use the mapping combined with lattice gauge theory to derive a gauge-invariant model of a system of three-level atoms. Finally, we prove that such system does not display photon condensation.

\subsection{No-go theorem for $M$-level systems}
\label{sec: No Go for m-level}

Before proceeding with the proof of the no-go theorem of a truncated model, we review the procedure to construct $M$-level models that are gauge-invariant, despite the Hilbert space truncation.

Recently, the generalized minimal coupling replacement, introduced in Ref.~\cite{DiStefano19} has been related to the general framework of lattice gauge theory and to the so-called Peierls substitution \cite{Savasta21}. Here, we show that also in the case of $M$-level systems this relationship remains valid. 
The Hamiltonian of any $M$-level system can be written in the basis of the eigenstates $\ket{m}$ as

\begin{equation}
\label{eq: m-level hamiltonian}
    \hat{h}_0 = \sum_{m = 1}^{M} \epsilon_m \dyad{m} ~.
\end{equation}

In the Coulomb gauge, and in the case of a single-mode
spatially uniform vector potential $A = A_0 (\hat{a} + \hat{a}^\dag)$,
such system can be coupled to $A$ as following \cite{DiStefano19}

\begin{equation}\label{NPg}
    \hat{h}_C = \hat{\mathcal{U}}_1 \hat{h}_0 \hat{\mathcal{U}}_1^\dag + \hat{\mathcal{H}}_{\rm ph}~,
\end{equation}
where $\hat{\mathcal{H}}_{\rm ph} =\hbar \omega_{\rm ph} \hat{a}^\dag \hat{a}$ is the cavity Hamiltonian, and $\hat{\mathcal{U}}_1 = \exp[-i e/(c \hbar) A_0 \hat{X} (\hat{a} + \hat{a}^\dag)]$ has the purpose of carrying out the minimal coupling replacement within the dipole approximation. Here $\hat{X} = \hat{P} \hat{x} \hat{P}$ (with $\hat{P} = \sum_{m = 1}^{M} \dyad{m}$) represents the truncated position operator.

We now consider a collection of $N$ identical, non-interacting $M$-level atoms. The total bare Hamiltonian is

\begin{equation*}
    \hat{\mathcal{H}}_0 = \sum_{n = 1}^N \sum_{m = 1}^M \epsilon_m \dyad{m_n}~,
\end{equation*}
and, by applying the method discussed above, we get the total interacting light-matter Hamiltonian:

\begin{equation}
    \label{eq: general interaction hamiltonian truncated}
    \hat{\mathcal{H}}_C = \hat{\mathcal{U}} \hat{\mathcal{H}}_0 \hat{\mathcal{U}}^\dag + \hat{\mathcal{H}}_{\rm ph}~,
\end{equation}
where $\hat{\mathcal{U}} = \exp[-i e/(c \hbar) A_0 \sum_n \hat{X}_n (\hat{a} + \hat{a}^\dag)]$, and $\hat{X}_n$ is the truncated position operator corresponding to the $n$-th atom.

We now show that, once the Hamiltonian of a generic $M$-level matter system interacting with an electromagnetic field has the structure in Eq.~\eqref{eq: general interaction hamiltonian truncated}, the photon annihilation operator $\hat{a}$ cannot have a non-vanishing expectation value in the ground state of this system. We demonstrate it using an approach similar to that adopted in Section~\ref{sect:general theorem}, based on the method developed in Ref.~\cite{Knight} for the standard minimal coupling replacement case.
We suppose that the ground state $\ket{\psi_0}$ of a system described by the Hamiltonian \eqref{eq: general interaction hamiltonian truncated} has the property that $\mel{\psi_0}{\hat{a}}{\psi_0} \neq 0$.
We introduce the following unitary operator:
\begin{equation}
    \label{eq: unitary op aharonov}
    \hat{\mathcal{T}} = \exp \left[ -i  \frac{e}{c \hbar} A_0 2\Re(\alpha) \sum_n \hat{X}_n  \right]\hat{\mathcal{D}}(\alpha)\,,
\end{equation}
which has the property of shifting the electron momentum and, in particular, to shift the photon operators 
\begin{eqnarray}
\label{eq:photonShift3}
\hat{\mathcal{T}} \hat{a} \hat{\mathcal{T}}^\dag = \hat{a} - \alpha~.
\end{eqnarray}

Again, we construct the trial state as $\ket{\psi} = \hat{\mathcal{T}}^\dag \ket{\psi_0}$, which is characterized by a zero order parameter,

\begin{equation*}
    \mel{\psi}{\hat{a}}{\psi} = 0~. 
\end{equation*}
Similarly to Sect.~\ref{sect:general theorem}, by means of Eq.~\eqref{eq:photonShift3}, we can prove that,
\begin{eqnarray}
    \label{eq: DUD2}
   \hat{\mathcal{D}}(\alpha) \hat{\mathcal{U}}^\dagger \hat{\mathcal{D}}^\dagger(\alpha)= 
   \exp \left[ -i  \frac{e}{c \hbar} A_0 2\Re(\alpha) \sum_n \hat{X}_n  \right]\hat{\mathcal{U}}^\dagger~,\nonumber\\
\end{eqnarray}
and following the steps of Sect.~\ref{sect:general theorem} we can prove Eq.~\eqref{eq: TU}, $\hat{\mathcal{U}}^\dagger\hat{\mathcal{T}}^\dagger=\hat{\mathcal{D}}^\dagger(\alpha)\hat{\mathcal{U}}^\dagger$, also for the present case. The energy of the trial state $\ket{\psi}$ reads,
\begin{eqnarray} 
    \label{eq: HcOverPsiIII}
 \braket{\psi|\hat{\mathcal{H}}_C|\psi}=  \braket{\psi_0|\Big[ \hat{\mathcal{T}}\hat{\mathcal{U}} \hat{\mathcal{H}}_0 \hat{\mathcal{U}}^\dag \hat{\mathcal{T}}^\dagger+\hat{\mathcal{T}}\hat{\mathcal{H}}_{\rm ph} \hat{\mathcal{T}}^\dagger \Big]|\psi_0}~.\nonumber\\
\end{eqnarray}
By means of Eq.~\eqref{eq: TU} and Eq.~\eqref{eq:photonShift3}, the energy reads,

\begin{eqnarray}
    \label{eq: no-go demonstration}
    \mel{\psi}{\hat{\mathcal{H}}_C}{\psi} &=& \mel{\psi_0}{\hat{\mathcal{H}}_C}{\psi_0} - \omega_{\rm ph} \abs{\alpha}^2 ~.
\end{eqnarray}

Eq.~\eqref{eq: no-go demonstration} implies that the state $\ket{\psi_0}$, which has a non-vanishing expectation value of the photon annihilation operator $\hat{a}$, is not the real ground state of the system, since a lower energy state $\ket{\psi}$, which was built specifically to have a vanishing  expectation  value,  has  a  lower  energy.
This ends the proof by contradiction. We have shown that the true ground-state of $\hat{\mathcal{H}}_C$ is characterized by a vanishing super-radiant order parameter $\braket{\hat{a}}$.
\subsection{Mapping onto a tight-binding lattice}
\label{app:LatticeBasis}

It has been shown that, in the dipole approximation, a two-level atom interacting with  the electromagnetic field  can be equivalently  described as a double-well system, where only the two lowest energy eigenstates are considered, which in turn corresponds to a two-site system interacting with a cavity field~\cite{Savasta21}. Here we extend this idea to a generic $M$-level system, showing that it can be mapped onto a linear chain of sites connected by hopping processes (i.e.~a tight-binding lattice).

We now define the following operator,
\begin{equation}
\label{eq:rOperator}
 \hat{R}=-\frac{e}{\hbar c}A_0 \hat{x}~.
\end{equation}
In the basis of the eigenstates $\ket{m}$, $ \hat{R}$ can be expressed as
\begin{equation}
\hat{ R} =  \sum^{M-1}_{m_1=0}\sum^{M-1}_{m_2=0} 
R_{m_1, m_2} \ket{m_1}\bra{m_2}~.
\end{equation}
Since $\hat{ R}$ is an Hermitian operator, it defines a basis of eigenvectors $\ket{r}$ such that:
\begin{equation}
\hat{ R} \ket{r} = \lambda_r  \ket{r}~.
\end{equation}
Recalling Eq.~\eqref{eq:rOperator}, the states $\ket{r}$ are also eigenvectors of the position operator, i.e.~$\hat{ x} \ket{r} = x_r  \ket{r}$, with $ \lambda_r=-e/(\hbar c)A_0 x_r$. As we will show momentarily, this local basis of eigenstates of the position operator $\hat{x}$ defines a natural lattice representation of the Hamiltonian $\hat h_0$.

We now introduce the unitary transformation $\hat O$, which connects the energy basis $\ket{m}$ with the position basis $\ket{r}$. Its matrix elements will be denoted by the symbol $O_{r,m} \equiv \bra{r} \ket{m}$. By definition, the following property holds true:
\begin{equation}\label{eq:Lambda}
\delta_{r_1,r_2} \lambda_{r_1}=
\sum^{M-1}_{m_1=0}\sum^{M-1}_{m_2=0}  O_{r_1, m_1}  R_{m_1, m_2} {O^\dagger}_{m_2, r_2}~.
\end{equation}
As this identity shows, the transformation $\hat O$ diagonalizes the position operator $\hat{ R}$.

The lattice representation of the matter Hamiltonian $\hat h_0$ is given by 
\begin{equation}\label{eq:h0MLV}
 \hat{ h}_0 = \sum^{M-1}_{r_1=0}\sum^{M-1}_{r_2=0} t_{r_1, r_2} \ket{r_1}\bra{r_2}~,
\end{equation}
where the hopping matrix $t_{r_1, r_2}$ is defined by
\begin{equation}
 t_{r_1, r_2} = \sum^{M-1}_{m=0}
 O_{r_1, m}  \epsilon_{m} {O^\dagger}_{m, r_2}~.
\end{equation}
It is worth noting that the Hamiltonian written above is on the same form of the one-body Hamiltonian on a lattice described by Eq.~\eqref{eq: one-body electron Hamiltonian}.

We are now in the position to write the Hamiltonian $\hat{h}_{\rm c}$ (defined by Eq.~\eqref{NPg}) in terms of the eigenvectors $\ket{m}$ of the position operator:
\begin{eqnarray}\label{eq:hMLV}
 \hat{h}&=&\sum^{M-1}_{r_1=0}\sum^{M-1}_{r_2=0}  e^{i \lambda_{r_1} (\hat{a}+\hat{a}^\dagger)} t_{r_1,r_2} e^{-i \lambda_{r_2} (\hat{a}+\hat{a}^\dagger)} \ket{r_1}\bra{r_2} \nonumber\\
 &=&
 \sum^{M-1}_{r_1=0}\sum^{M-1}_{r_2=0}  e^{-i (e/\hbar c)  (x_{r_1}-x_{r_2}) (\hat{a}+\hat{a}^\dagger)} t_{r_1,r_2} \ket{r_1}\bra{r_2} 
 ~.\nonumber\\
\end{eqnarray}
This is the main result of this Section. It shows that the coupled Hamiltonian $\hat{h}_{\rm c}$ has the exact same form of a tight-binding lattice model coupled to light via the Peierls substitution. 
Actually, the Peierls method was developed to study electron systems interacting with static magnetic fields, in the framework of the tight-binding approximation. The Peierls substitution can be regarded as an anticipation of lattice gauge theory, which is the general method developed by Wilson for studying non-perturbative relativistic gauge theories on a lattice~\cite{Wilson}, or in condensed matter physics, to analyze quantum simulations of lattice gauge theories~\cite{Wiese}. Here we have shown that the two methods coincide provided that one operates in the position basis $\ket{m}$.
Hence, in the lattice basis, the Peierls substitution is the most general tool to couple matter with a single cavity mode.

\subsection{Example: ladder three-level system}
\label{subsect:TLA}

We now consider the particular case of a three-level ladder atom, which can be described as a three-site system with inversion symmetry, as depicted in Fig.~\ref{fig:three level}. 
In this Section we show that, in stark contrast to the conclusions of Refs.~\cite{Hayn11,Baksic13}, such system does not display photon condensation.

The bare Hamiltonian of a single three-level ladder atom, expressed in the lattice representation (see Eq.~\eqref{eq:h0MLV}), reads as following:
\begin{equation}
\hat h_{0} = \sum_{i=-1}^{1} \epsilon_i \vert i \rangle \langle i \vert + t (\vert -1 \rangle \langle 0\vert + \vert 0 \rangle \langle 1\vert + {\rm H. c.})~.
\end{equation}
We consider here a system with parity symmetry, so that the selection rules for a three-level ladder atom apply: $\epsilon_{-1}= \epsilon_1$. From now on, we also fix $\epsilon_{-1}= \epsilon_1 = 0$.
According to gauge lattice theory, the interaction with the electromagnetic field can be obtained by introducing the Wilson parallel transporter \cite{Wilson}.  The resulting Hamiltonian, after applying the dipole approximation (uniform field), is
\be \label{H}
\hat h_{\rm tot}= \hat{\mathcal{ H}}_{\rm ph} + {\hat h}\, ,
\ee
where $\hat H_{\rm ph}$ is the free-photon Hamiltonian 
and ${\hat h}$ is the atomic Hamiltonian, now invariant under arbitrary (site-dependent) phase transformations:
\begin{equation}
    \hat h =  \epsilon_0 \vert 0 \rangle \langle 0\vert 
   + [ t e^{-i \gamma (\hat a^\dag + \hat a) }(\vert -1 \rangle \langle 0\vert + \vert 0 \rangle \langle 1\vert) + {\rm h. c.}]~,
\label{Hp}
\end{equation}
accordingly to Eq.~\eqref{eq:hMLV}.
Here, $\gamma = -e d A_0/(\hbar c)$ with  $d$ the distance between two adjacent sites.
For simplicity, we assume a single mode optical resonator: $\hat H_{\rm ph} = \hbar \omega_{\rm ph} \hat a^\dag \hat a$, with the field coordinate $\hat A = A_0 (\hat a^\dag + \hat a)$, where $A_0$ is the vacuum fluctuation amplitude.

The Hamiltonian in Eq.~\eqref{Hp} can also be written as
\begin{equation}\label{H'0U}
  \hat h = \hat {\cal U}_1   \hat h_0 \hat {\cal U}_1^\dag\, ,
\end{equation}
where 
\begin{equation}
\hat  {\cal U}_1 = \exp \left[-i \frac{e}{\hbar c} \hat x_L \hat A\right]\, ,
\end{equation}
and  $\hat x_L$ is the lattice coordinate, i.e.~$\hat x_L = d\sum_j j \vert j \rangle \langle j\vert$.

Let us now consider a collection of ${N}$ identical, non-interacting three-level ladder atoms. The total Hamiltonian is 
\be\label{HN}
\hat{ \mathcal{H} }= \hat{ \mathcal{H} }_{\rm ph} + 
\epsilon_0\hat{\Sigma}_{0,0} 
   +t [ e^{-i \gamma (\hat a^\dag + \hat a) }(\hat{\Sigma}_{-1,0} + \hat{\Sigma}_{0,1}) + {\rm h. c.}]
\, ,
\ee
where 
\be
\hat{\Sigma}_{i,j} = \sum_{k=1}^{N} \vert i_k \rangle \langle j_k\vert~.
\ee
Equation \eqref{HN} can be written compactly as
\be\label{HNU}
\hat{ \mathcal{H} } = \hat{ \mathcal{H} }_{\rm ph} + 
\hat {\cal U} \hat H_0 \hat {\cal U}^\dag
\, ,
\ee
where 
\be
\hat{ \mathcal{H} }_0 = 
\epsilon_0 \hat{\Sigma}_{0,0} + t \left(
\hat{\Sigma}_{-1,0} + \hat{\Sigma}_{0,1} + {\rm h.c.} \right)
\ee
and
\be
\hat {\cal U} = \exp \left[ i \frac{\gamma}{2} \left( \hat{a} + \hat{a}^\dag \right) \sum_{j=1}^{N} j\, \hat{\Sigma}_{j,j} \right] \, .
\ee

When the system's Hamiltonian is cast in the form of  Eq.~\eqref{HNU}, the theorem demonstrated in Subsection~\ref{sec: No Go for m-level}, showing that no photon condensation can occur, can be readily applied to this case.



%
\section{Summary and conclusions}
\label{sect:summary_and_conclusions}

Previous no-go theorems have demonstrated that gauge invariance forbids phase transitions to a photon condensate state when the cavity-photon mode is assumed to be {\it spatially uniform}. In this Article, we have shown that a matter system in the absence of magnetic interactions, even if interacting with a non-uniform electric field, cannot display photon condensation. This is in agreement with the findings of Refs.~\cite{Basko19,Andolina20,guerci} where it was shown that photon condensation can occur in a spatially-varying cavity field, where the magnetic field interacts with the electronic system, and that it is formally equivalent to a magneto-static instability.

The actual theoretical description of realistic electron systems requires unavoidable approximations. It has been shown that such approximations can spoil gauge invariance. Since no-go theorems are  related to gauge invariance, it is possible that these approximate models yield superradiant phase transitions, which would not be allowed when considering the full infinite-dimensional Hilbert space. For example, it has been theoretically predicted that a collection of three-level systems coupled to light can display a first-order phase transition to a photon condensate state~\cite{Hayn11,Baksic13}. Recently, it has been shown that the gauge principle can be formulated in a consistent way also when considering matter systems described in truncated Hilbert spaces. In this Article, we have shown that the no-go theorem forbidding spontaneous photon condensation in the ground state remains valid for these approximate models satisfying gauge invariance.  In particular, we have presented a non-perturbative no-go theorem for: (i) systems of interacting electrons roaming both a continuous space and on a lattice; (ii) an ensemble of non-interacting $M$-level systems. We also discussed the case of three-level ladder atoms, showing that if this system is described by a model satisfying the gauge-invariance principle (in the truncated space) no photon condensation can occur, in agreement with the general theorem and in stark contrast to the conclusions reached by the authors of Refs.~\cite{Hayn11,Baksic13}.

We note that the conclusions reached in this Article apply to light-matter interacting systems whose interaction is described by the minimal coupling replacement. However, our conclusions have to be carefully reconsidered when applied to systems like superconducting artificial atoms coupled to microwave resonators, since these do not display the coordinate-momentum interaction resulting from the minimal coupling replacement~\cite{nataf_naturecommun_2010}. One may argue that, at a microscopic level, also these systems interact according to the gauge-invariance principle and hence via the minimal (or Zeeman) coupling replacement~\cite{viehmann_prl_2011}. However, we observe that, in several circuit-QED systems, artificial atoms interact with the electromagnetic resonator through a magnetic flux. Our no-go theorem naturally does not apply to this class of systems, where a magnetic field is present. 

Finally, we would like to mention that the lattice gauge theory approach employed here can be fruitfully applied to condensed matter lattice models, such as the Hubbard model~\cite{Hubbard63}, the Falikov-Kimball model~\cite{EFKM} and even more complicated multi-orbital systems. While our general theorem holds irrespective of all the microscopic details, it would be interesting to study strongly interacting systems in the presence of a cavity magnetic field, transcending the hypothesis of our theorem.
For such an investigation, it is crucial to use a correctly gauge invariant model, which can be obtained with the methods of lattice gauge theory.
Lattice gauge theory is in general necessary to correctly describe quantum materials strongly coupled to light, not only in the context of photon condensation~\cite{Schiro20,Basko19,mazza_prl_2019,andolina_prb_2019}, but also in studying other phenomena, such as cavity-induced ferroelectricity~\cite{Ashida20}, light-induced topological properties \cite{Tokatly20,Sentef19}, photon-mediated superconductivity \cite{Schlawin_prl_2019,curtis_prl_2019,allocca_prb_2019,Ebbesen19}, and photo-chemistry \cite{Rubio2018,Rubio2018a,Rubio2019,Rubio2020,Koch2020,Koch2021}.

\section{Acknowledgements}

M.P. was supported by the European Union's Horizon 2020 research and innovation programme under grant agreement no.~881603 - GrapheneCore3 and by the MUR - Italian Ministry of University and Research - under the ``Research projects of relevant national interest  - PRIN 2020"  - grant no. 2020JLZ52N, project title: ``Light-matter interactions and the collective behavior of quantum 2D materials (q-LIMA)''. F.M.D.P. was supported by the Universit\`a degli
Studi di Catania, Piano di Incentivi per la Ricerca di
Ateneo 2020/2022, progetto QUAPHENE and progetto Q-ICT.
S.S. acknowledges the Army Research Office (ARO)
(Grant No. W911NF1910065).

\appendix

\section{\label{app: Derivation dressed electron fields}Derivation of the dressed electron field}
One of the starting point of this work was to establish the correct and gauge-invariant method to couple light to matter. A system of interacting electrons, described by the electron field $\hat{\psi}({\bm r})$, can be coupled to a single-mode electromagnetic field by using Eq.~\eqref{eq: dressing electronic fields}, which we write again
\be
\hat {\cal U} \hat \psi ({\bm r})\hat {\cal U}^\dag = e^{i({ e/c\hbar}) \chi({\bm r})(\hat a + \hat a^\dag)} \hat \psi ({\bm r})\, .
\ee
The proof of this formula is straightforward, ad it follows from the fermionic anti-commutation property of the electron field $\{ \hat{\psi}({\bm r}), \hat{\psi}^\dag({\bm r'}) \} = \delta(\bm{r} - \bm{r}')$ and from the Baker-Campbell-Hausdorff formula, that is
\be
\label{eq: (appendix) Campbell-Hausdorff formula}
e^{\hat{S}} \hat{O} e^{-\hat{S}} = \hat{O} + \comm{\hat{S}}{\hat{O}} + \frac{1}{2!} \comm{\hat{S}}{\comm{\hat{S}}{\hat{O}}} + \dots \, ,
\ee

where is our case 
\begin{eqnarray}
 \hat{S} &=& -i { \frac{e}{c\hbar}} (\hat a + \hat a^\dag) \int d {\bm r'}\, \hat \psi^\dag({\bm r'})
\chi({\bm r'}) \hat \psi({\bm r'})~,\\
\hat{O}& =& \hat{\psi}({\bm r'})~.
\end{eqnarray}
Thus, the commutator $[\hat{S}, \hat{O}]$ becomes
\bea
\comm{\hat{S}}{\hat{O}} &=& -i { \frac{e}{c\hbar}} (\hat a + \hat a^\dag) \comm{\int d {\bm r'}\, \hat \psi^\dag({\bm r'})
\chi({\bm r'}) \hat \psi({\bm r'})}{\hat{\psi}({\bm r})} \nonumber \\
&=& -i { \frac{e}{c\hbar}} (\hat a + \hat a^\dag) \int d {\bm r'}\, \comm{ \hat \psi^\dag({\bm r'}) \hat \psi({\bm r'})
}{\hat{\psi}({\bm r})} \chi({\bm r'}) \nonumber \\
&=& i { \frac{e}{c\hbar}} (\hat a + \hat a^\dag) \int d {\bm r'}\, \delta({\bm r'} - {\bm r}) \hat{\psi}({\bm r'})  \chi({\bm r}) \nonumber \\
&=& i { \frac{e}{c\hbar}} (\hat a + \hat a^\dag) \chi({\bm r}) \hat{\psi}({\bm r})
\eea
and, by recursively replacing it into Eq.~\eqref{eq: (appendix) Campbell-Hausdorff formula}, we obtain the dressed electron field.

The procedure expressed above is also valid in the case of a lattice, whose space domain is no more continuous, and it can be obtained simply substituting $\int d{\bm r} \to \sum_{\bm r_i}$ and $\hat{\psi}({\bm r}) \to \hat{c}_{\bm r_i}$.

\section{No-Go theorem for a multi-mode cavity field}
\label{app:multi-mode}
The proof of the no-go theorem for photon condensation discussed in Sect. \ref{sect:general theoremA} can be extended also to the case of a multi-modes cavity, characterized by $N$ photonic modes. The multi-mode photonic Hamiltonian reads,
\begin{eqnarray}
    \label{eq: Hph}
 \hat{H}_{\rm ph}=\sum_{i=1}^N\hbar \omega_{{\rm ph},i}\hat{a}^\dagger_i\hat{a}_i ~.
\end{eqnarray}
We denote as $\ket{\psi_0}$ the super-radiant ground state. In the multi-mode scenario, we take as superradiant order parameter the following quantity,
\begin{eqnarray}
\label{eq:orderp}
 \sum_{i=1}^N|\alpha_i|^2 ~,
\end{eqnarray}
where $\alpha_i\equiv\braket{\psi_0|\hat{a}_i|\psi_0}$. Since we are assuming that $\ket{\psi_0}$ is a super-radiant ground state, at least one of $\alpha_i$ should be non-zero.
The light-matter coupling is implemented via the following unitary,
\be 
\label{eq:transf s}
\hat {\cal U} = \exp\left[
-i { \frac{e}{c\hbar}}\sum^N_{i=1}  (\hat a_i + \hat a_i^\dag)
\int d {\bm r}\, \hat \psi^\dag({\bm r})
\chi_i({\bm r}) \hat \psi({\bm r})
\right]~.
\ee
The electronic field transforms as,
\be
\label{eq: dressing electronic fields s}
\hat {\cal U} \hat \psi ({\bm r})\hat {\cal U}^\dag = e^{i({ e/c\hbar})\sum\limits^N_{i=1} \chi_i({\bm r})(\hat a_i + \hat a_i^\dag)} \hat \psi ({\bm r})\, 
\ee

Following the procedure described in Sect.~\ref{sect:general theorem} we now consider the following generalized multi-mode unitary operator,
\begin{equation}
    \label{eq: sec_quant no-go unitary op s}
    \hat{\mathcal{T}} = \exp \left[ -i \frac{2 e}{c \hbar} \sum_{i=1}^N \Re (\alpha_i) \int d {\bm r} \hat{\psi}^\dag ({\bm r}) \chi_i({\bm r}) \hat{\psi} ({\bm r}) \right] \prod_i\hat{\mathcal{D}}_i(\alpha_i)   ~,
\end{equation}
where $\hat{\mathcal{D}}_i(\alpha_i)=\exp{- \alpha_i^* \hat{a}_i + \alpha_i \hat{a}_i^\dag} $ is the displacement operator for the $i$-th characterized by a displacement $\alpha_i$.  
The electronic and photonic fields transform under $\hat{\mathcal{T}}$ as,
\begin{eqnarray}
\label{eq: shift of nogo oper 1s}
    \hat{\mathcal{T}} \hat{\psi} ({\bm r}) \hat{\mathcal{T}}^\dag &=& e^{i \frac{2 e}{c \hbar} \sum_i\Re (\alpha_i) \chi_i({\bm r})} \hat{\psi}({\bm r})
\\
\label{eq:shiftaApp}
\hat{\mathcal{T}} \hat{a}_i \hat{\mathcal{T}}^\dag &=& \hat{a}_i - \alpha_i~.
\end{eqnarray}
The trial state $\ket{\psi}$ is constructed by applying the operator $ \hat{\mathcal{T}}^\dagger$ to the super-radiant ground state $\ket{\psi_0}$, i.e. $\ket{\psi}=\hat{\mathcal{T}}^\dagger\ket{\psi_0}$. By means of Eq.~\eqref{eq:shiftaApp} we have $\braket{\psi|\hat{a}_i|\psi}=0$ for $i=1,..,N$.
In the multi-mode case, we can generalize Eq.~\eqref{eq: DUD}, by means of Eq.~\eqref{eq:shiftaApp} obtaining that,
\begin{align}
    \label{eq: DUDs}
   \hat{\mathcal{D}}_i(\alpha_i)& \hat{\mathcal{U}}^\dagger \hat{\mathcal{D}}_i(\alpha_i)^\dag=\nonumber\\&=\exp \left[- i \frac{2 e}{c \hbar} \Re (\alpha_i) \int d {\bm r} \hat{\psi}^\dag ({\bm r}) \chi_i({\bm r}) \hat{\psi} ({\bm r}) \right]  \hat{\mathcal{U}}^\dagger~.\nonumber\\
\end{align}
We now consider the product $\hat{\mathcal{U}}^\dagger\hat{\mathcal{T}}^\dagger$, by using Eq.~\eqref{eq: DUDs} we have,
\begin{eqnarray}
    \label{eq: TUapp}
\hat{\mathcal{U}}^\dagger\hat{\mathcal{T}}^\dagger=\prod_i\hat{\mathcal{D}}^\dagger_i(\alpha_i) \hat{\mathcal{U}}^\dagger~.
\end{eqnarray}
Finally, we calculate the total energy of the trial state $\ket{\psi}=\hat{\mathcal{T}}^\dagger\ket{\psi_0}$,
\begin{eqnarray} 
    \label{eq: HcOverPsiIIIApp}
 \braket{\psi|\hat{\mathcal{H}}_C|\psi}=  \braket{\psi_0|\Big[ \hat{\mathcal{T}}\hat{\mathcal{U}} \hat{\mathcal{H}}_0 \hat{\mathcal{U}}^\dag \hat{\mathcal{T}}^\dagger+\hat{\mathcal{T}}\hat{\mathcal{H}}_{\rm ph} \hat{\mathcal{T}}^\dagger \Big]|\psi_0}~.\nonumber\\
\end{eqnarray}
By using the property in Eq~.\eqref{eq: TUapp}, and by means of Eq.~\eqref{eq:shiftaApp}, Eq.~\eqref{eq: HcOverPsiIIIApp} can be cast as,
\begin{eqnarray}
    \label{eq: HcOverPsiApp}
 \braket{\psi|\hat{H}_C|\psi}=\braket{\psi_0|\hat{H}_C|\psi_0}  -\sum_i\hbar \omega_{{\rm ph},i}|\alpha_i|^2 ~.
\end{eqnarray}
Since we are assuming that $\psi_0$ is a super-radiant ground state, there exist at least one $i$ such that $\alpha_i\neq0$. Hence,
\begin{eqnarray}
    \label{eq: HcOverPsiApp2}
 \braket{\psi|\hat{H}_C|\psi}<\braket{\psi_0|\hat{H}_C|\psi_0} ~.
\end{eqnarray}
We have found a state $\ket{\psi}$ which has a lower energy that $\ket{\psi_0}$. This is clearly in contrast with the initial hypothesis that $\psi_0$ is the ground state. Hence, this conclude the proof by contradiction.




\end{document}